\documentclass{article} % For LaTeX2e
\usepackage{iclr2023_conference,times}

\usepackage{amsmath,amsfonts,bm}

\def\eqref#1{equation~\ref{#1}}
\def\1{\bm{1}}

\DeclareMathAlphabet{\mathsfit}{\encodingdefault}{\sfdefault}{m}{sl}
\SetMathAlphabet{\mathsfit}{bold}{\encodingdefault}{\sfdefault}{bx}{n}

\usepackage{hyperref}
\usepackage{url}
\usepackage{graphicx}
\usepackage{subcaption}
\usepackage[inkscapeformat=png]{svg}

\title{DiffESM: Conditional Emulation of Earth System Models with Diffusion Models}

\author{Seth Bassetti$^1$, Brian Hutchinson$^{1,2}$, Claudia Tebaldi$^3$, Ben Kravitz$^4$\\
$^1$ Computer Science Department, Western Washington University, Bellingham, WA\\
$^2$ Foundational Data Science Group, Pacific Northwest National Laboratory, Richland, WA\\
$^3$ Joint Global Change Research Institute, Pacific Northwest National Laboratory, College Park, MD\\
$^4$ Earth and Atmospheric Sciences Department, Indiana University, Bloomington, IN\\
\texttt{\{bassets,brian.hutchinson\}@wwu.edu, claudia.tebaldi@pnnl.gov, bkravitz@iu.edu}
}

\iclrfinalcopy % Uncomment for camera-ready version, but NOT for submission.
\begin{document}

\maketitle

\begin{abstract}
    Earth System Models (ESMs) are essential tools for understanding the impact of human actions on Earth's climate. One key application of these models is studying extreme weather events, such as heat waves or dry spells, which have significant socioeconomic and environmental consequences. However, the computational demands of running a sufficient number of simulations to analyze the risks are often prohibitive. In this paper we demonstrate that diffusion models -- a class of generative deep learning models -- can effectively emulate the spatio-temporal trends of ESMs under previously unseen climate scenarios, while only requiring a small fraction of the computational resources. We present a diffusion model that is conditioned on monthly averages of temperature or precipitation on a $96 \times 96$ global grid, and produces daily values that are both realistic and consistent with those averages. Our results show that the output from our diffusion model closely matches the spatio-temporal behavior of the ESM it emulates in terms of the frequency of phenomena such as heat waves, dry spells, or rainfall intensity. 
\end{abstract}

\section{Introduction}
Earth System Models (ESMs) play an important role in estimating the risk of extreme weather events under different emissions scenarios. The rarity of such weather events means that data must be aggregated over numerous runs to get reliable statistics. However, the computational demands of ESMs limits the number of realizations that can be performed. By using existing data to learn the statistical characteristics of ESM output, emulators can address this issue, by generating thousands of realizations on the scale of minutes or hours rather than weeks or months. Machine learning approaches are well-suited to building such emulators, especially generative deep learning methods capable of learning to approximate complicated, high dimensional distributions. We present a denoising diffusion probabilistic model that learns to closely model the spatio-temporal behavior of an ESM, producing month-long samples of either daily mean temperature or precipitation. 
Our emulator, DiffESM, can be steered to generate samples under novel climate scenarios (or existing climate scenarios for which we want to enlarge the sample size of daily variables) by conditioning generation on a monthly mean map of the climate variable. 
Such monthly mean maps can be produced by existing emulators, like fldgen~\citep{link2019fldgenv1} or STITCHES~\citep{tebaldi2022stitches}.
Once trained, the emulator offers a dramatic improvement over traditional ESMs in terms of speed, allowing for rapid investigation of the effect of climate scenarios on the distribution of extreme weather events, making it a valuable tool for climate researchers and policy-makers. 

Many researchers have utilized machine learning for weather and climate modeling.
One key application is the use of machine learning for forecasting, including now-casting \citep{nowcasting-radar, nowcasting-precip}, sub-seasonal forecasting \citep{Sub-seasonal-forecasting, sub-season-ensemble} and seasonal climate forecasting \citep{seasonal-forecast-sa}. Using machine learning to improve the resolution of ESMs is another active area of research, as many ESM outputs are too coarse for local-scale predictions. Data-driven methods can be used to construct models of local-scale phenomena; for example, modeling clouds with physics-informed neural networks \citep{NeuralNetworkConservationEnergy, rasp2018deep},
or improving the resolution of regional climate models with generative adversarial networks (GANs) and --- more recently --- diffusion models \citep{addison2022machine, precip-fields, stengel2020adversarial}. Certainly, computationally efficient emulation of ESMs themselves has a long tradition, from statistics-based methods \citep{stats-emulation, stats-emulation-2} to generative deep learning approaches \citep{ayala2021loosely, ayalaconditional, puchko2020deepclimgan}. In contrast to these prior generative deep learning approaches to ESM emulation using GANs, which are notoriously difficult to train, we find that our DiffESM is significantly easier to train and better approximates the ESMs as measured by several of the same metrics.

\section{Methods}

Diffusion models generate samples from a target distribution via an iterative denoising process that maps samples from a known distribution (Gaussian) to samples from the unknown target distribution. This iterative denoising is often easier to accomplish than attempting to map noise samples to the target distribution directly. Such models are trained by progressively destroying the information in real samples using a forward process, and then learning to progressively reconstruct the destroyed sample \citep{denoising-diffusion, improved-diffusion}. 

Our model architecture is highly inspired by the Video Diffusion \citep{video-diffusion} and Imagen Video~\citep{imagen-video} model architectures.
Specifically, for each denoising step we use a fully convolutional U-Net \citep{u-net} architecture with interleaved spatial and temporal convolution layers. We exclude self-attention due to computational limitations. The input to the model is a noisy sample of shape $C \times T \times H \times W$, where $C=1$ is the number of variables (temperature or precipitation), $T=28$ is the sequence length in days, and $H=96$ and $W=96$ are the spatial dimensions of the grid. The model outputs a sample of the same size, which represents daily temperature or precipitation values for each spatial location over the globe for a 28-day (i.e., four week) ``month.'' The architecture consists of four downsampling/upsampling layers with a bottleneck layer in between. Each layer uses two ResNet blocks \citep{ResNet}, a temporal-only convolution operation, and a respective upsampling or downsampling convolutional operation. The bottleneck has the same structure except for the lack of upsampling or downsampling. In addition to downsampling or upsampling the spatial dimension in each layer, the model increases the channel dimension at each depth. The respective channel dimensions per level are: 48, 128, 192, and 256 respectively. To steer the outputs of our model, we provide as conditioning: A spatial map of the monthly average of the variable, the day of the year that the 28-day sequence begins on, and the timestep that indicates the stage of the reverse (denoising) diffusion process.

\begin{figure}
  \subcaptionbox*{Monthly Hot Streak}[.33\linewidth]{%
    \includegraphics[width=\linewidth]{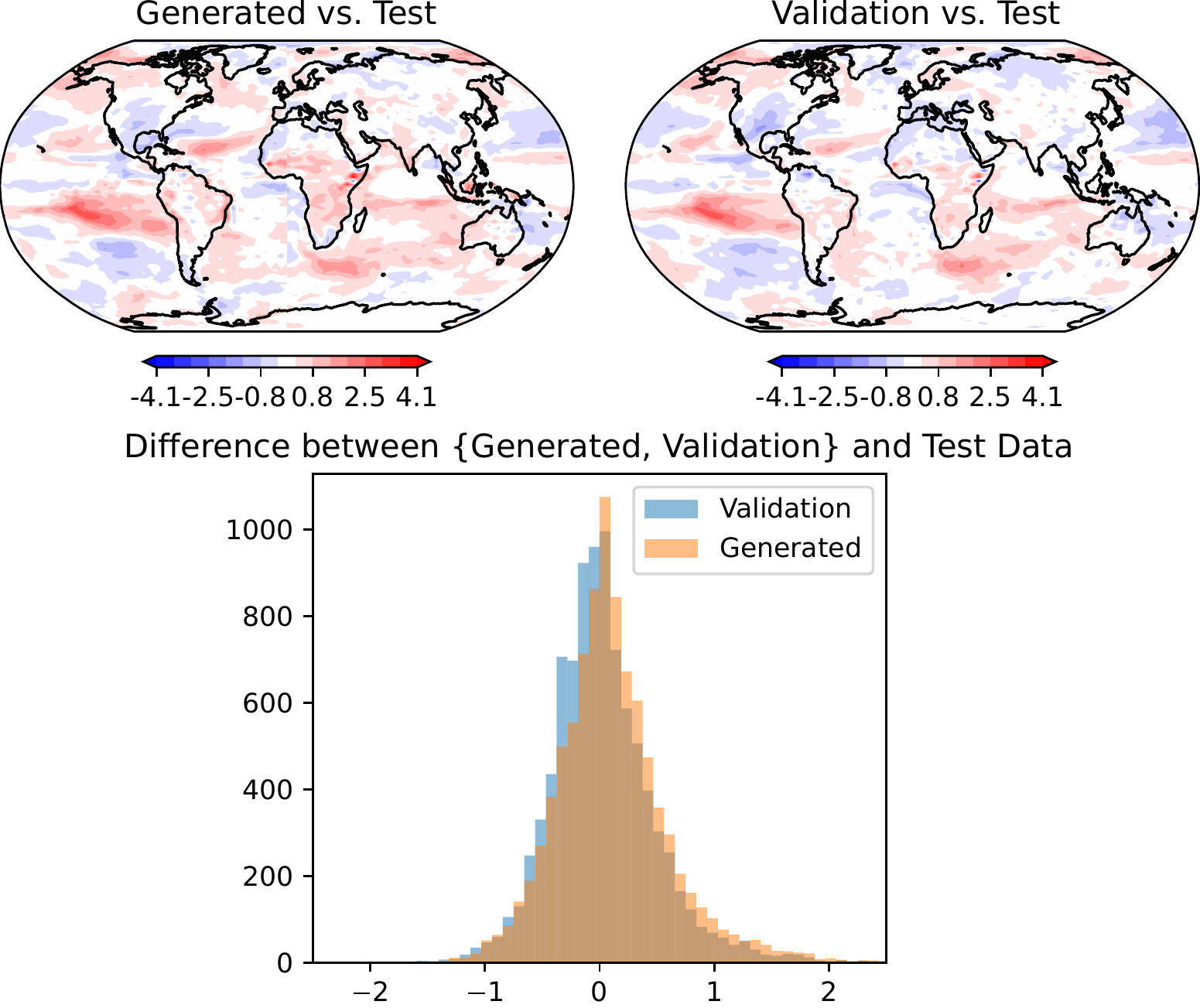}%
  }%
  \hfill
  \subcaptionbox*{Monthly Hot Days}[.33\linewidth]{%
    \includegraphics[width=\linewidth]{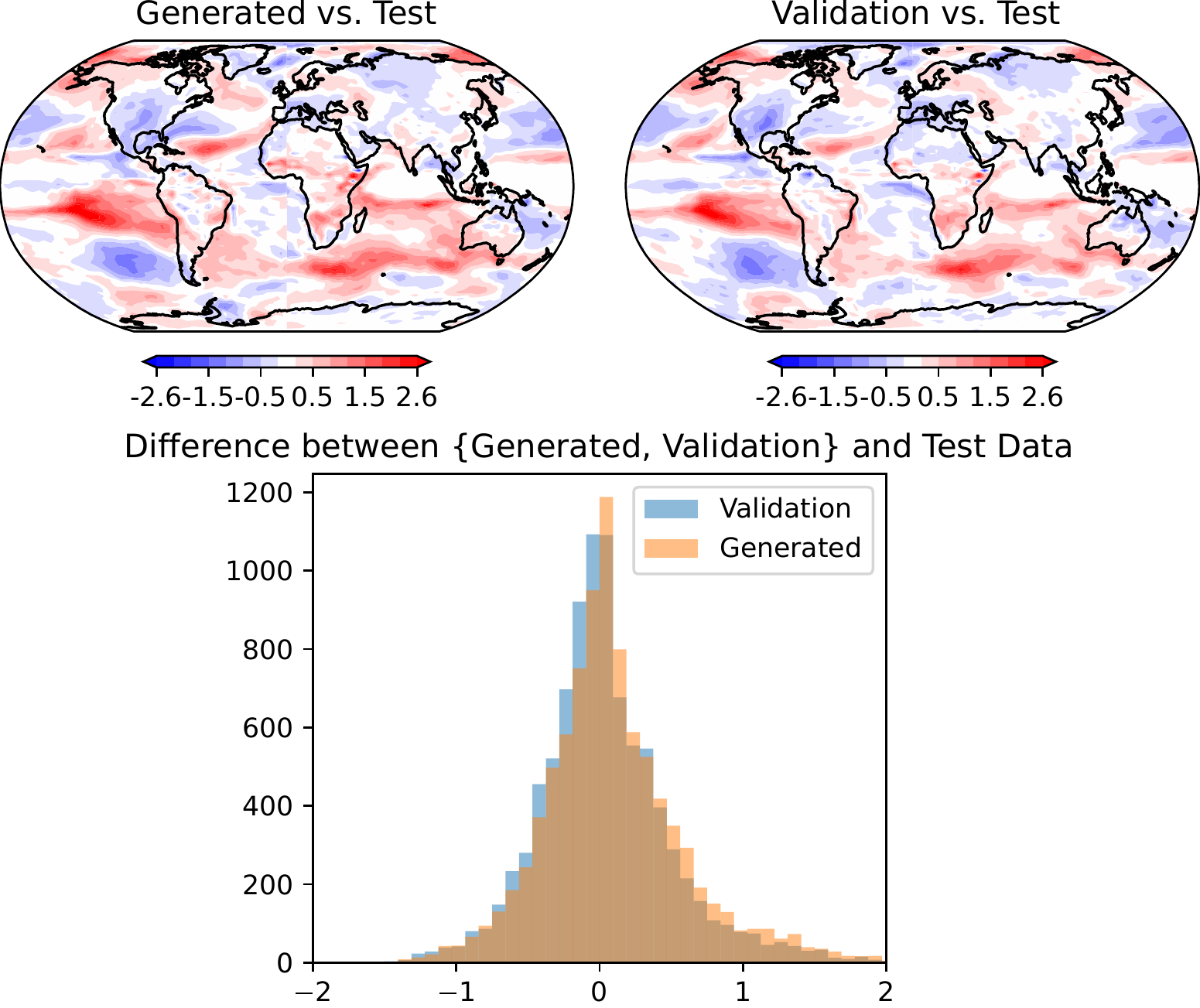}%
  }%
  \hfill
  \subcaptionbox*{90th Quantile Values}[.33\linewidth]{%
    \includegraphics[width=\linewidth]{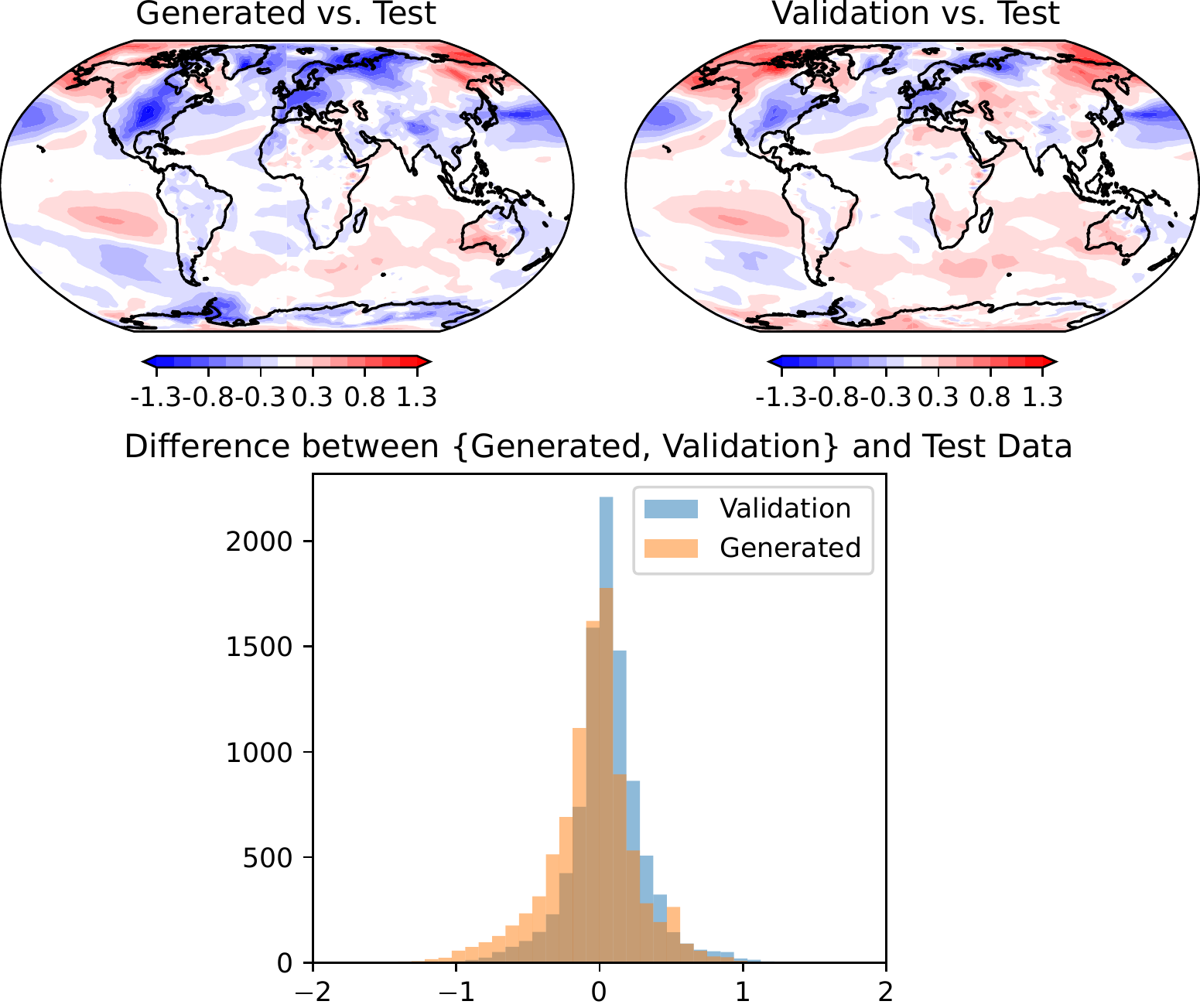}%
  }%
  \hfill
  \subcaptionbox*{Monthly Dry Spell}[.33\linewidth]{%
    \includegraphics[width=\linewidth]{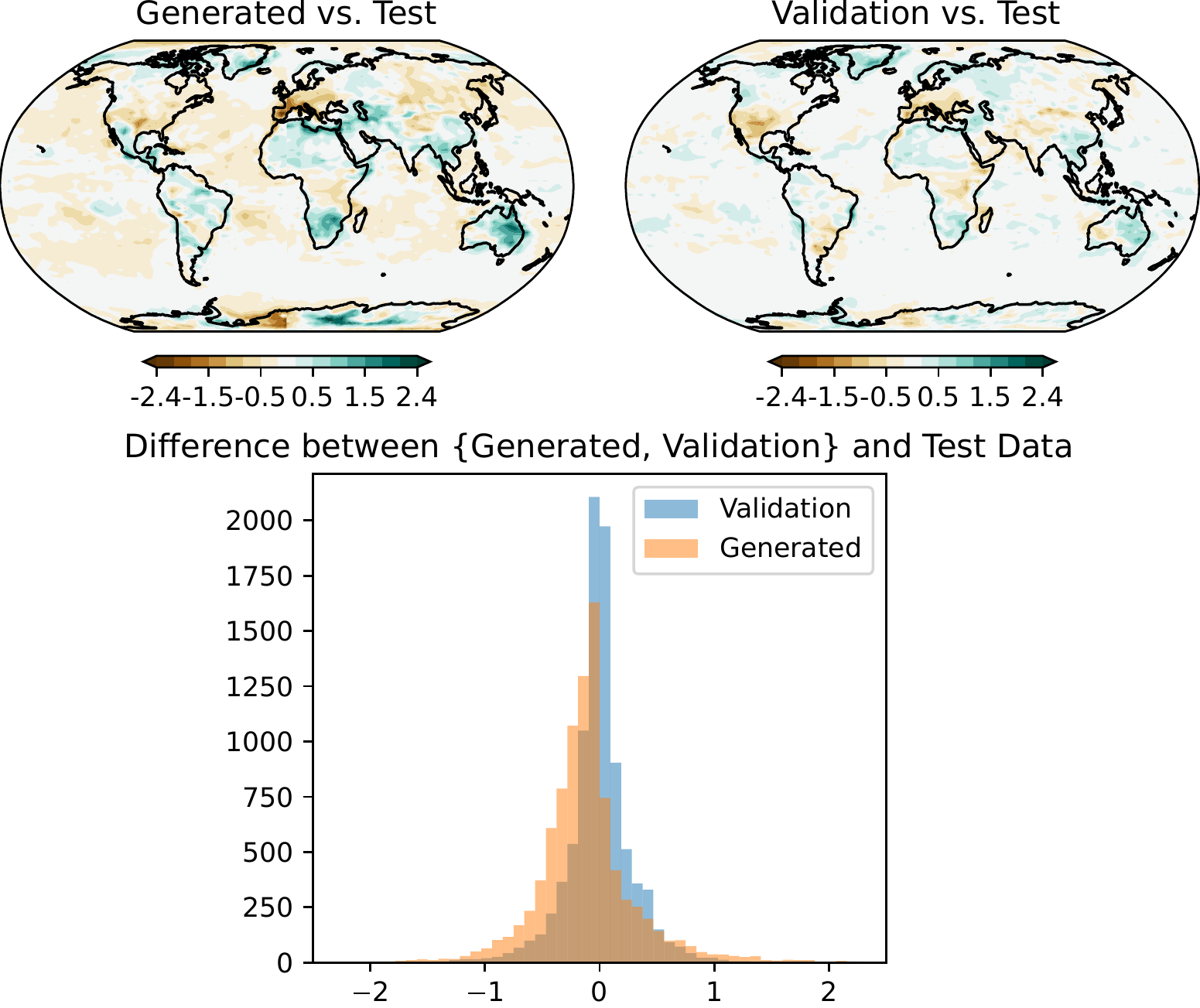}%
  }%
  \hfill
  \subcaptionbox*{Monthly Dry Days}[.33\linewidth]{%
    \includegraphics[width=\linewidth]{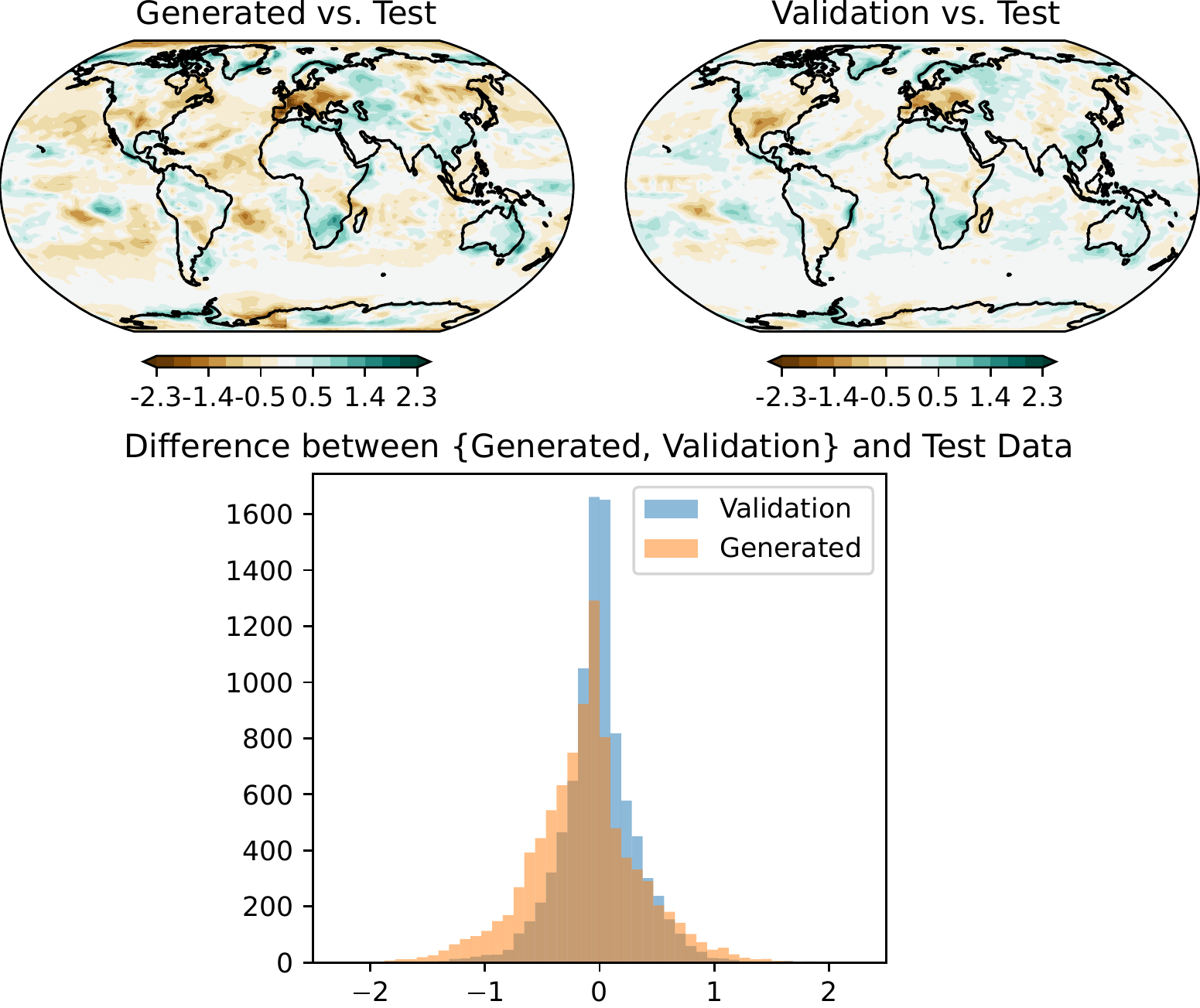}%
  }%
  \hfill
  \subcaptionbox*{SDII (Wet Day Precipitation)}[.33\linewidth]{%
    \includegraphics[width=\linewidth]{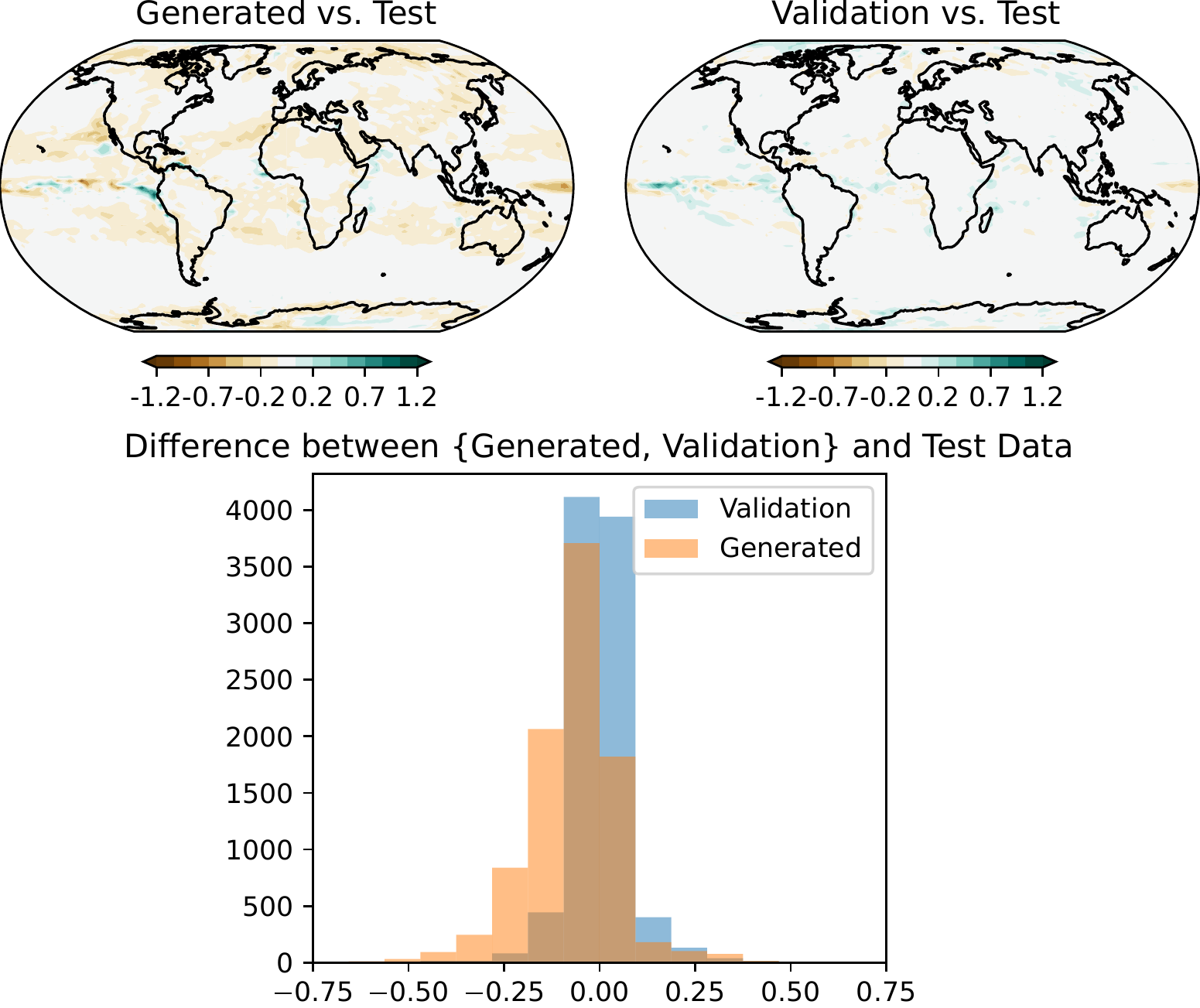}%
  }%
  \caption{Error Histograms and Spatial Maps Averaged across 2080 - 2100} \label{fig:maps}
\end{figure}

To train, we use a continuous-time diffusion model \citep{variational-diffusion} with v-parameterization \citep{v-param}. The timesteps for noising each sample are randomly chosen from $(0,1]$. The noisy samples, along with all conditioning, are passed into the model, and we use mean squared error to impose a reconstruction loss on the v-term and the model's outputs. We additionally implement classifier-free guidance on the day labels \citep{classifier-free-guidance}. During training, 15\% of the day labels are randomly dropped out, letting the model learn a joint unconditional and conditional representation of the data.
Each model is trained for a total of 10 epochs with a batch size of 256 split between four GPUs. We use the Adam optimizer \citep{Adam} with a learning rate of 0.0004, and $\beta_1$ and $\beta_2$ initialized to 0.9 and 0.99. All sampling is done with 250 timesteps, uniformly spaced between 0 and 1.

\section{Experiments}

\subsection{Dataset}
Our dataset is composed of daily output on a $96 \times 96$ spatial grid from the IPSL-CM5A ESM. In this study, we use only the daily mean temperature and daily precipitation variables. Our dataset consists of six total realizations (initial condition ensemble members), each representing the outputs from a full run of the ESM from pre-industrial times to 2100. These realizations are split into a ``historical'' period of values from 1850 to 2006 and realizations from the RCP8.5 scenario from 2006 to 2100. A ``scenario'' represents a potential human-driven emission pathway, and RCP8.5 represents the most extremescenario, in the sense of projecting the highest, unmitigated greenhouse gas emissions among the set of scenarios that are run by climate models to explore the range of uncertain future anthropogenic forcings on the climate system. 
We use four realizations of combined historical and RCP8.5 data to form a training set, one realization of historical and RCP8.5 as a validation set, and one realization as a test set. To test the model's generalization capabilities to novel climate scenarios, we use one realization each from RCP4.5, a less extreme emission scenario not seen during training, as distinct validation and test sets. For brevity, in this paper we will report results on cross-scenario performance in the 2080-2100 period, as the matched-scenario and earlier future tasks are easier.

\subsection{Metrics}
The goal of our emulator is not to accurately predict a single realization of the future climate, but rather, model realizations from a spatio-temporal statistical distribution that closely correlates with the ESM's distribution. Towards this end, we look at the spatial and temporal distributions of statistics computed from each 28-day ``month,'' such as the number of days in the month exceeding the 90th percentile temperature, the length of the longest dryspell in the month, or the average precipitation on days exceeding the 90th percentile. 
First, for each of the 252 validation set months between 2080-2100 (inclusive), we create a monthly mean map by averaging over the 28 days, and then generate one 28-day sample from our diffusion model.
We then compute statistics at each spatial location over each monthly sample in the validation, test, and generated sets. For each statistic, we then average over all 252 months, giving us one spatial map per dataset. We produce two signed difference maps: validation minus test (which differ only due to internal variability produced by the ESM) and generated minus test, showing the similarity in spatial distribution between the two pairs of datasets.
We do not expect either of these difference maps to be exactly zero, due to inherent variability between runs of the same ESM, and hope only that the level of variability between generated and test is comparable to the level of variability between validation and test.

\subsection{Results}
Figure~\ref{fig:maps} plots six pairs of the difference maps described in the previous section, with three temperature statistics in the top row and three precipitation statistics in the bottom row. For each pair, the gen-test map is on the left while the val-test map is on the right. Below each pair of maps we also show overlaid histograms of the $9216=96^2$ spatial difference values (orange from gen-test and blue from val-test). The maps show strong correlation in the differences produced by the generated and validation sets. The histograms show that the distribution of differences for gen-test is very close to that of val-test, especially for the temperature metrics. 
We do note a tendency for the generated data to slightly under-predict rainfall values.

\begin{figure}
  \subcaptionbox*{}[.5\linewidth]{%
    \includegraphics[width=\linewidth]{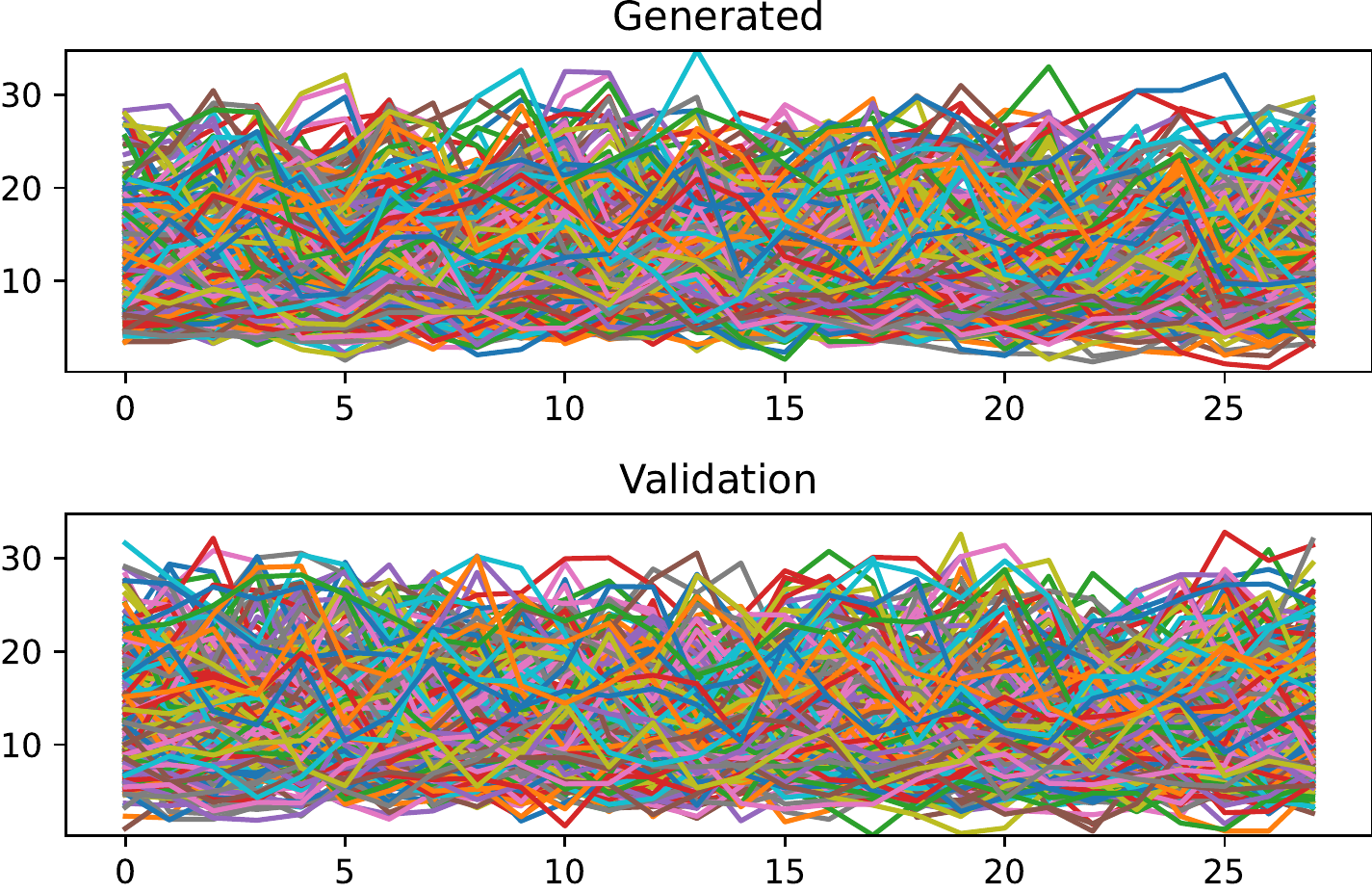}%
    \caption{28-day overlaid temperature sequences}
  }%
  \hfill
  \subcaptionbox*{}[.5\linewidth]{%
    \includegraphics[width=\linewidth]{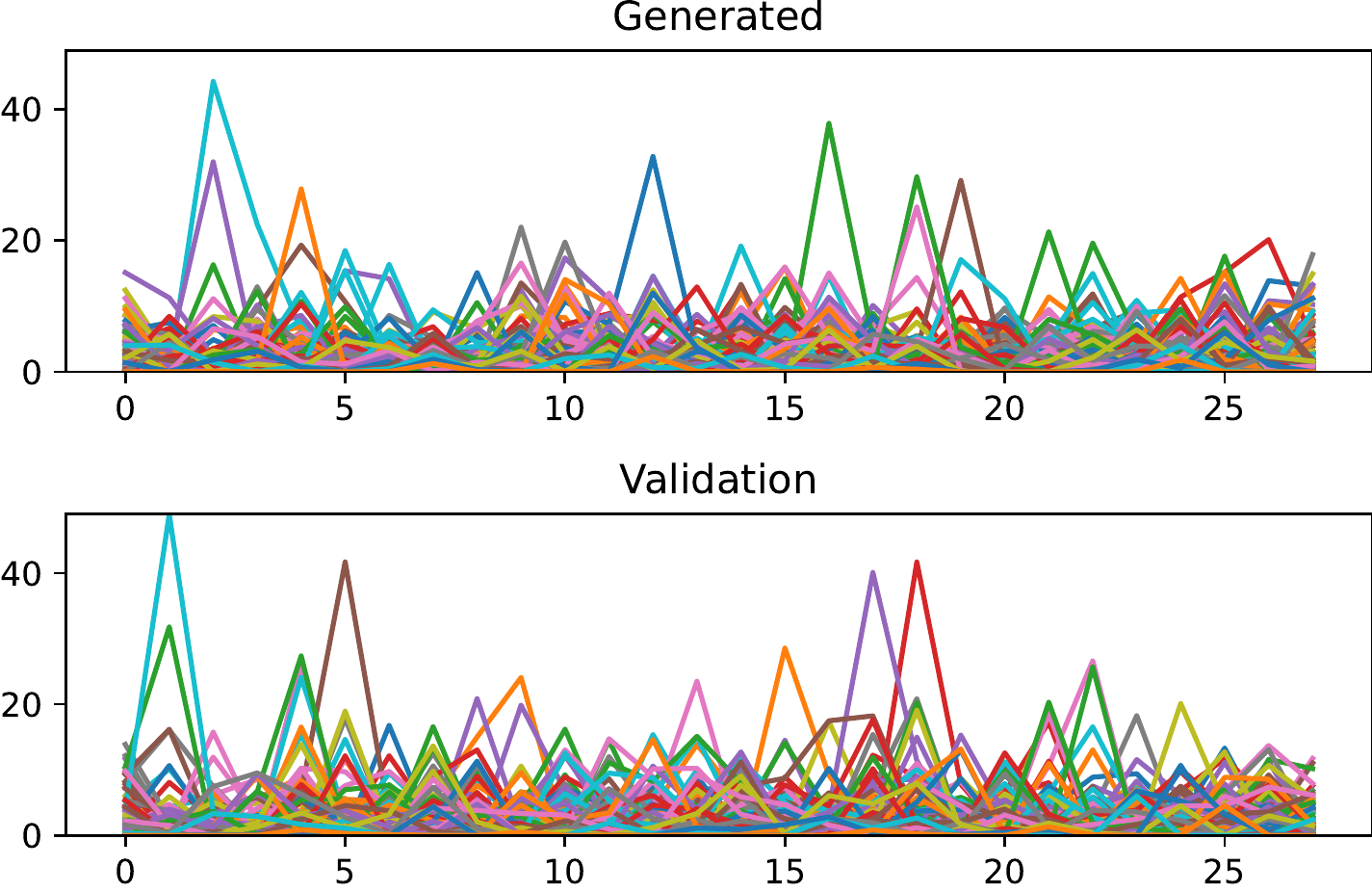}%
    \caption{28 day overlaid precipitation sequences}
  }
  \caption{Distribution of 28-day sequences of daily values for 252 months from 2080-2100 in Melbourne, Australia.} \label{fig:samples}
\end{figure}

As qualitative examples of the time series produced by our model, the 252 validation and 252 generated samples used to compute the above results were also plotted in Fig.~\ref{fig:samples} as temperature or precipitation time series at the spatial location closest to Melbourne, Australia. Each line represents a 28 day sequence from the years 2080-2100, all overlaid on top of each other. They show, qualitatively, that the temporal behavior of our samples approximates that of the validation data.

\section{Conclusion and Future Work}
In this paper, we have demonstrated the capability of conditional video diffusion models to emulate ESM output of daily temperature and precipitation under a climate scenario unseen during training.
We observe that the samples produced by our models are comparable to those of ESMs in several extreme-relevant metrics, such as frequency and spatial distribution of hot streaks or dry spells, and intensity of precipitation during extremely wet days. The ability to generate such simulations in a timely manner will significantly enhance our ability to characterize the risks of extreme weather events under various future climate scenarios. Another -- pragmatic -- use of emulation of daily quantities from monthly means could be as a solution to decrease the cost of archiving and handling ESM output, which is becoming increasingly high due to ESMs' higher and higher resolution. 

There are numerous directions for future work. One promising area would be to integrate multiple variables into a single diffusion model, since modeling the correlation between temperature and precipitation would likely lead to increased performance. This would also result in output that preserves the joint characteristics of the variables and allow to address more consistently those types of extremes that result from the combination of hot and dry, or cool and wet behavior of the climate system. Despite its speed advantages over ESMs, the diffusion models could themselves be further sped up using sampling techniques such as progressive distillation \citep{v-param}. Lastly, while the work reported in this paper emulates just one ESM and evaluates on one novel scenario, we plan to replicate these findings over multiple ESMs and scenarios to provide further evidence of the promise of these techniques.

\subsubsection*{Acknowledgments}
This work was conducted with the support of the US Department of Energy, Office of Science, as part of the GCIMS project within the MultiSector Dynamics program area of the Earth and Environmental System Modeling program. The authors also thank the NVIDIA corporation for the donation of GPUs used in this work. 

\bibliography{refs}
\bibliographystyle{iclr2023_conference}

\end{document}